\documentstyle[aps,prd,preprint,epsfig]{revtex}

\begin{document}
\draft
\title{O(4) texture with a cosmological constant}

\author{Inyong Cho\footnote{Electronic address: 
inyong@nuclecu.unam.mx}}
\address{Instituto de Ciencias Nucleares,
Universidad Nacional Autonoma de Mexico, \\
Apartado Postal: 70-543,
Mexico D.F. 04510, Mexico}

\date{\today}

\maketitle

\begin{abstract}
We investigate O(4) textures in a background with a positive
cosmological constant.
We find static solutions which co-move with the expanding background.
There exists a solution in which the scalar field is regular at the horizon.
This solution has a noninteger winding number smaller than one.
There also exist solutions in which scalar-field derivatives are 
singular at the horizon.
Such solutions can complete one winding within the horizon.
If the winding number is larger than some critical value,
static solutions including the regular one 
are unstable under perturbations.
\end{abstract}

\vspace{0.5in}
\pacs{PACS numbers: 11.27.+d, 98.80.Cq, 04.40.-b}

\section{Introduction}
Phase transitions associated with the symmetry breaking in the early
universe could produce topological defects. 
Those are core defects such as domain walls, cosmic strings, and 
monopoles, or textures~\cite{Vilenkin}.

Global textures are produced when a continuous global symmetry 
$G$ is broken to a group $H$.
The resulting homotopy group $\pi_D(G/H)$ is nontrivial,
and the vacuum manifold is $S^D$; for $G$=O(N) models, 
$H$=O(N-1) and $D=N-1$.
The mapping from the physical space to the vacuum manifold is
$R^D \to S^D$. 

The Lagrangian describing such texture models is
\begin{equation}
{\cal L} = -{1\over 2}\partial_\mu\Phi^a\partial^\mu\Phi^a
-{\lambda\over 4}(\Phi^a\Phi^a -\eta^2)^2\,,
\label{eq=Lag}
\end{equation}
where $a=1,...,N$.
This Lagrangian has O(N) symmetry. 
After the phase transition, the scalar field acquires
the value in the vacuum manifold, $\Phi^a\Phi^a =\eta^2$,
of which symmetry is O(N-1).

The energy of this model in $D$ spatial dimensions is given by
\begin{equation}
E=\int d^Dx \left[ {1\over 2}(\nabla\Phi^a)^2 + V(\Phi^a)\right]
\equiv I_1 +I_2\,.
\label{eq=E}
\end{equation}
By transforming $\Phi^a(x)$ under the rescaling $x\to\alpha x$,
the energy configuration becomes
\begin{equation}
E_\alpha = \alpha^{2-D}I_1 +\alpha^{-D}I_2\,.
\label{eq=Ealpha}
\end{equation}
For a global texture, 
the potential part of the energy vanishes, $I_2=0$, since
the scalar field sits on the vacuum manifold in the entire
space $R^D$.
Equation~(\ref{eq=Ealpha}) tells us that the energy 
is a function of the rescaling factor $\alpha$.
The configuration relaxes to acquire the minimum energy.
Therefore, O(2) texture ($D=1$) expands ($\alpha$ decreasing), 
and O(4) texture ($D=3$) collapses
since there is no stable minima in $E_\alpha$ 
with respect to (w.r.t.) $\alpha$.
Only O(3) texture ($D=2$) is stable.
This is the well-known Derrick's theorem~\cite{Derrick}.

Collapsing O(4) textures produced during the Grand Unified
phase transitions attracted much interest in the early 1990's 
due to their potential cosmological significance~\cite{Turok,Davis,Vachaspati,Notzold}. 
As the universe expands and cools down,
the texture knots shrink and radiate away into Goldstone bosons. 
The texture collapse occurs at the speed of light, and
correlates the scalar field on larger and larger scales. 
Quickly the correlation length grows
to the horizon scale, and expands with it.
As the texture collapses, the gradient energy of the scalar field
gets concentrated at the center of the texture.  
This energy becomes large enough for the scalar field to overcome
the potential barrier, and to leave the vacuum manifold. 
The texture unwinds in the end.
During the collapsing texture leaves a signature in the structure formation
of the universe.

Textures do not require a complete winding for collapse.
Above some critical noninteger winding number $\sim 0.5$,
textures collapse~\cite{Perivolaropoulos,Sornborger}.
The stability of textures in an expanding background has also been
investigated for the radiation-dominated, matter-dominated and 
inflation era~\cite{Sornborger,Chen,Barabash}. 

A non-vanishing cosmological constant has attracted
increasing  attention
as a source for an accelerating universe.
Supernova observation data show that the universe is accelerating
rather than decelerating. 
The simplest source for such expansion is a cosmological constant.

If it existed, the story of texture stability could be altered.
With a cosmological constant the universe undergoes de Sitter expansion.
O(4) textures formed in such a background could slow down their collapse,
or could even stabilize themselves to co-move with the background.

In this paper, we investigate O(4) textures with a cosmological constant.
We study both the scalar and the gravitational field configurations.
We solve the field equations numerically, 
and seek the static solutions.
We also check the stability of them.

When there is a positive cosmological constant, a horizon develops
in the geometry. 
There exists a static texture solution which is regular across the horizon.
However, the scalar field of this solution does not complete its winding
beyond the horizon; the winding number is nonintegral $\sim 0.88$.
There also exist static solutions in which scalar-field derivatives are
singular at the horizon; however, the scalar field itself is ``finite''
at the horizon. 
The value of the scalar field at the horizon can be 
either higher than that of the regular solution, or lower.
For this type of solution, the texture
can even complete one winding within the horizon. 
Both of the regular and the singular solutions are physical.

The regular solution is unstable under a small perturbation.
The singular solutions of the high scalar field at the horizon are 
also unstable. The others look stable.

In Sec.~II, we present the model and field equations.
In Sec.~III, we study texture solutions in pure de Sitter space.
In Sec.~IV, we include the texture self-gravity. 
In Sec.~V, we check the stability of the solutions.
In Sec.~VI, we discuss the texture evolution.
We conclude in Sec.~VII.

\section{The model}
The action describing an O(4) texture with a cosmological constant is
\begin{equation}
{\cal S} = \int d^4x\sqrt{-g}
\left[ {{\cal R}-2\Lambda \over 16\pi G}
-{1\over 2}\partial_\mu\Phi^a \partial^\mu\Phi^a
-{\lambda\over 4}(\Phi^a\Phi^a -\eta^2)^2 \right]\,,
\label{eq=action}
\end{equation}
where $\Phi^a$ is the scalar field with $a=1,...,4$,
$\eta$ is the symmetry breaking scale,
and $\Lambda$ is the cosmological constant.
We are mainly interested in the case of $\Lambda > 0$.
We assume the spherical symmetry and the general static form of the
metric is
\begin{equation}
ds^2 = -B(r)dt^2 +A(r)dr^2 +r^2d\Omega_2^2\,.
\label{eq=metric}
\end{equation}

We consider a scalar field everywhere in space to reside in the
vacuum manifold, $\Phi^a\Phi^a=\eta^2$. 
The scalar field is of the form,
\begin{equation}
\Phi^a = \eta (\sin\chi \sin\theta \cos\phi,
\sin\chi \sin\theta \sin\phi,
\sin\chi \cos\theta, \cos\chi)\,,
\end{equation}
where $\theta$ and $\phi$ are the usual spherical coordinates.
$\chi$ is a function of the radial coordinate, $\chi (r)$.
In this case, the nonlinear $\sigma$-model approximation is applied,
and $\lambda$ in the potential is treated as a Lagrangian multiplier.
The scalar field equation, then, is
\begin{equation}
\nabla^\mu\partial_\mu\Phi^a = 
-{\Phi^b( \nabla^\mu\partial_\mu\Phi^b) \over \eta^2}\Phi^a
={(\partial^\mu\Phi^b)(\partial_\mu\Phi^b) \over \eta^2}\Phi^a\,.
\end{equation}

The energy-momentum tensor is given by
\begin{equation}
T_{\mu\nu} = \partial_\mu\Phi^a \partial_\nu\Phi^a
-{1\over 2}g_{\mu\nu}\partial_\alpha\Phi^a\partial^\alpha\Phi^a\,.
\end{equation}
With the given metric~(\ref{eq=metric}), the Einstein's equation leads to
\begin{eqnarray}
-G^t_t &=& {1\over A}\left( {A'\over Ar}+ {A-1\over r^2}\right)
	=8\pi G\eta^2 \left( {\chi'^2\over 2A}+{\sin^2\chi \over r^2}\right)
	+\Lambda\,,\\
-G^r_r &=& {1\over A}\left( -{B'\over Br}+ {A-1\over r^2}\right)
	=8\pi G\eta^2 \left( -{\chi'^2\over 2A}+{\sin^2\chi \over r^2}\right)
	+\Lambda\,,\\
-G^{\theta_i}_{\theta_i} &=&
{1\over A}\left[ -{B'' \over 2B}+{1\over 4}\left({B'\over B}\right)^2
+{A'B'\over 4AB}-{B'\over 2Br} + {A'\over 2Ar} \right]
	=8\pi G\eta^2 {\chi'^2\over 2A} +\Lambda\,,
\end{eqnarray}
and the scalar field equation leads to
\begin{equation}
{\chi''\over A} +{1\over A}\left({B'\over 2B}-{A'\over 2A}+{2\over r}
\right)\chi' -{\sin 2\chi \over r^2}=0\,.
\label{eq=scalar}
\end{equation}
In the field equations, 
$\Lambda$ can be absorbed into the radial coordinate by rescaling.
Then, the equations are free of $\Lambda$, and the solutions are as well.
This implies that the solutions look self-similar one another for different
$\Lambda$ scales. The only physical difference is that the configuration
stretches along the radial direction by the $\Lambda$ scale.

\section{Solutions in pure de Sitter space}
In this section, we investigate texture configurations
in pure de Sitter space without including the
self-gravity of the texture field.
In this case, solutions are more tractable
analytically as well as numerically.
The background geometry is
\begin{equation}
ds^2 = -(1-{\Lambda \over 3}r^2)dt^2 
+{1\over 1-{\Lambda \over 3}r^2}dr^2 +r^2d\Omega_2^2\,.
\end{equation}
The scalar field equation with this metric is
\begin{equation}
\chi'' +{2(1-2R^2) \over R(1-R^2)}\chi' - 
{\sin 2\chi \over R^2(1-R^2)} =0\,.
\label{eq=scalards}
\end{equation}
Here, we rescaled the radial coordinate by $R=(\Lambda /3)^{1/2}r$,
and the prime denotes differentiation w.r.t. $R$.
The horizon is located at $R=1$.
This equation is solved only numerically.
The numerical calculation is impossible at the horizon,
$R=1$, where the equation becomes singular.
However, if we fix the value of the scalar field at the horizon
as a boundary condition,
we can avoid the difficulty.
The equation is numerically solved by the relaxation method 
after fixing two-point boundary values at $R=0$ and $R=1$.
We locate the center of texture at the origin, $\chi (0)=0$.

\begin{figure}
\begin{center}
\epsfig{file=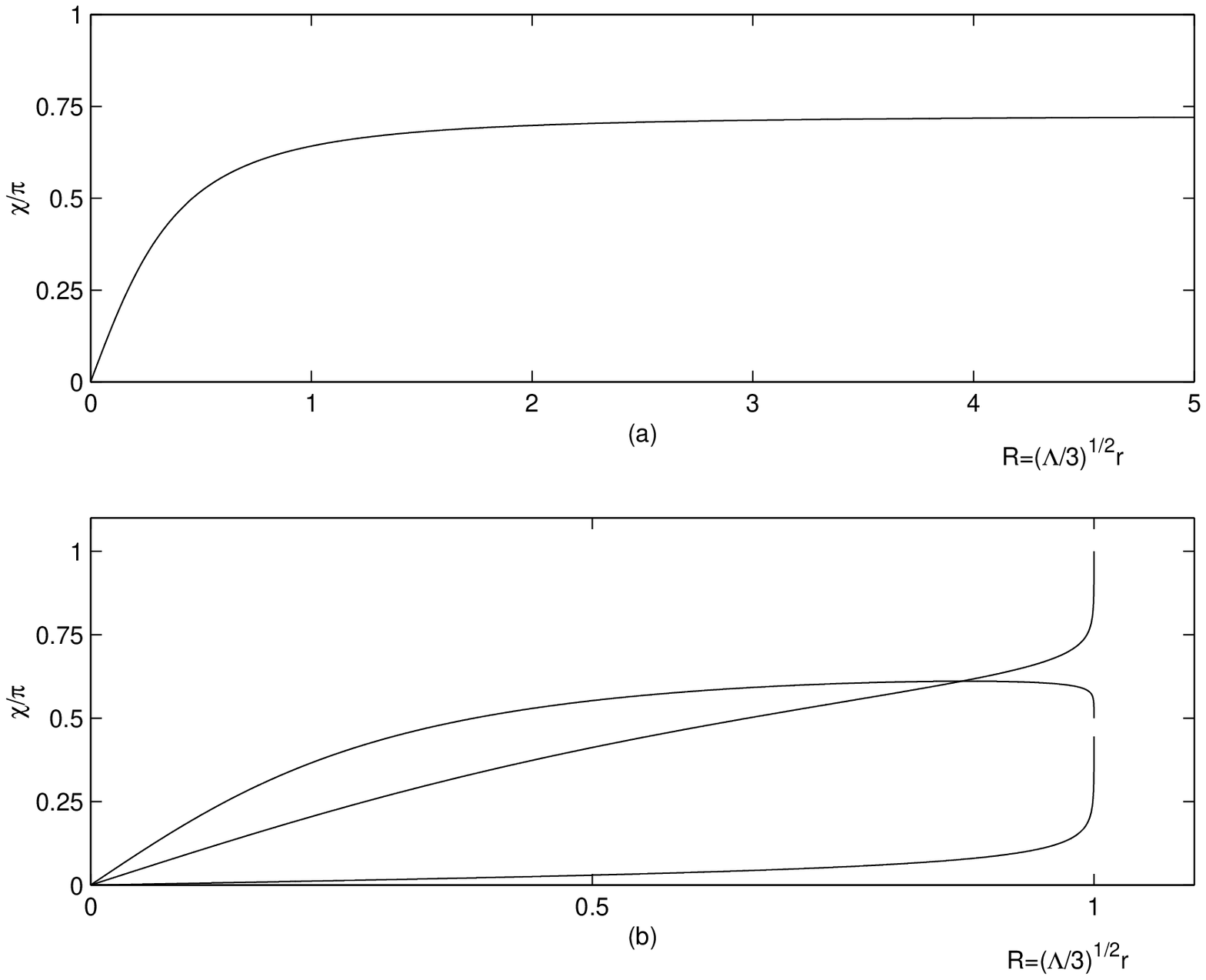,width=5in}
\end{center}
\vspace{0.5in}
\caption{
Plot of the scalar field in pure de Sitter space.
(a) Regular solution. At the horizon the scalar field has
$\chi_h \simeq 0.64192\pi$. It approaches asymptotically
$\chi \approx 0.72081\pi$.
(b) Singular solutions for $\chi_h=\pi$ and $\chi_h =0.445\pi$.
We also show an unconventional solution for 
$\chi_h =0.5\pi$ in which the scalar field decreases close to the horizon.}
\label{fig=dssfd}
\end{figure}

First, let us discuss the regular solution.
If $\chi$ is finite up to the second derivative at the horizon,
from Eq.~(\ref{eq=scalards}) we have a regularity condition for $\chi$
at the horizon as
\begin{equation}
2\chi'_h +\sin 2\chi_h = 0\,,
\label{eq=regds}
\end{equation}
where the subscript {\it h} denotes the value at the horizon.
We found numerically that this condition is satisfied when 
$\chi_h \simeq 0.64192\pi$.
If we proceed numerical calculation outside the horizon with
this value, the scalar field reaches 
$\chi (R\to\infty) \approx 0.72081\pi$ asymptotically.
The scalar field of this regular configuration 
does not complete winding in the vacuum manifold
even at infinity. The solution is plotted in Fig.\ref{fig=dssfd}(a).

Besides the regular solution, there exist solutions in which
$\chi''$ at the horizon is infinite.
From Eq.~(\ref{eq=scalards}), when $\chi''$ is infinite, unfortunately
there would not be a definite information of $\chi$ and $\chi'$
at the horizon; $\chi'$ can be either finite, or infinite
at the horizon. However, we observe that $\chi'$ approaches
a very large value close to the horizon. This indicates that
$\chi'$ may be infinite there---we checked that at least
$\chi'' \propto (R-1)^s$ type of solution does not
exist, where $s<0$.

Such solutions exist for $\chi_h \lesssim 0.445\pi$
and $0.65\pi \lesssim  \chi_h \lesssim 1.1\pi$.
However, close to $\chi_h \sim 0 $ and $\sim 0.65\pi$ 
the derivative
$\chi'_h$ is not very large though.
Other values of $\chi_h$ than those for regular and singular
solutions, field configurations are not so conventional; they do
not monotonically increase, but decrease close to the horizon.
The singular solutions are plotted in Fig.\ref{fig=dssfd}(b).

Note that in flat spacetime as well as in self-gravitating 
spacetime without a cosmological constant, 
there do not exist static configurations
above some critical noninteger winding number smaller than one.
However, with a cosmological constant, we have static solutions
up to $\chi_h \sim 1.1\pi$.
For $\chi_h =\pi$, the texture completes one winding at the horizon,
and for $\pi < \chi_h \lesssim 1.1\pi$, it does even inside the horizon.
We will see that this is also true for the case including the self-gravity
of the texture field in the next section. 
This result agrees with the expectation we mentioned in the Introduction.
Collapse is balanced by the expansion of the background.

For a given singular configuration inside the horizon, 
the matching solution outside must exist. However,
as the field equation~(\ref{eq=scalards}) does not provide any 
information about the functional form, or specific values
of the field derivatives  at the horizon,
we cannot continue numerical calculations to the region outside 
of the horizon.

Considering solutions only inside the horizon is justified as following.
When the texture is formed, the scalar field acquires values in the vacuum 
manifold. Different regions in space take different values of the field.
The correlation length scale within which the region takes the same value,
cannot exceed the cosmological horizon scale at the moment of formation.
Usually the formation of one defect is completed within the cosmological horizon.
Other defects are formed in the neighborhood.

In the present case, if we include the contribution of other matters 
(for example, the texture field itself) 
to the energy density of the universe,
the cosmological horizon determined by the total energy density
is always smaller than the de Sitter horizon.
Therefore, considering solutions only inside the de Sitter horizon is 
perfectly guaranteed.

Although singular solutions would be accepted by the above justification,
we need to comment some trouble associated with them.
First, if $\chi'$ is infinite at the horizon, the curvature
diverges. If we include the self-gravity of the texture,
the horizon is not only a coordinate singularity, but becomes 
a curvature singularity. 
If there is a curvature singularity, 
the signal emitted from it can propagate
to the inner region. 
Observers off the center of the texture eventually will see the
signal.
This trouble may be resolved if we put the causal boundary
to the cosmological horizon which is inside the de Sitter horizon.
Hence, the texture configuration is cut by neighboring textures
before it develops a singularity.

As we will see in Sec.~V, the static texture configurations are
unstable under small perturbations. The scalar field may evolve
in such a way as to remove the singular behavior at the horizon.

In addition, when the field gradient is very large locally, 
the scalar field
acquires enough energy to overcome the potential barrier.
Therefore, the nonlinear $\sigma$-model approximation is not valid any longer.

Even though the solutions obtained here are static inside the horizon,
they co-move with the expanding background.
The configuration will be self-similar as time passes unless 
there are perturbations.

\section{Self-gravitating solutions}
In this section, we include the effect of the texture self-gravity.
For textures of GUT symmetry-breaking scale, 
$\eta \sim 10^{-3}m_p \sim 10^{16}\text{GeV}$,
the effect is very small.
Therefore, the results are very similar to those in the previous section.

The location of the horizon is displaced a little bit from that of
pure de Sitter space.
Numerically we can perform calculations close to the horizon
where $1/A\approx 10^{-(4\sim 5)}$.
Beyond that point, the values in calculation exceed the numerical limit.
Therefore, we obtain numerical solutions up to that point.

\begin{figure}
\begin{center}
\epsfig{file=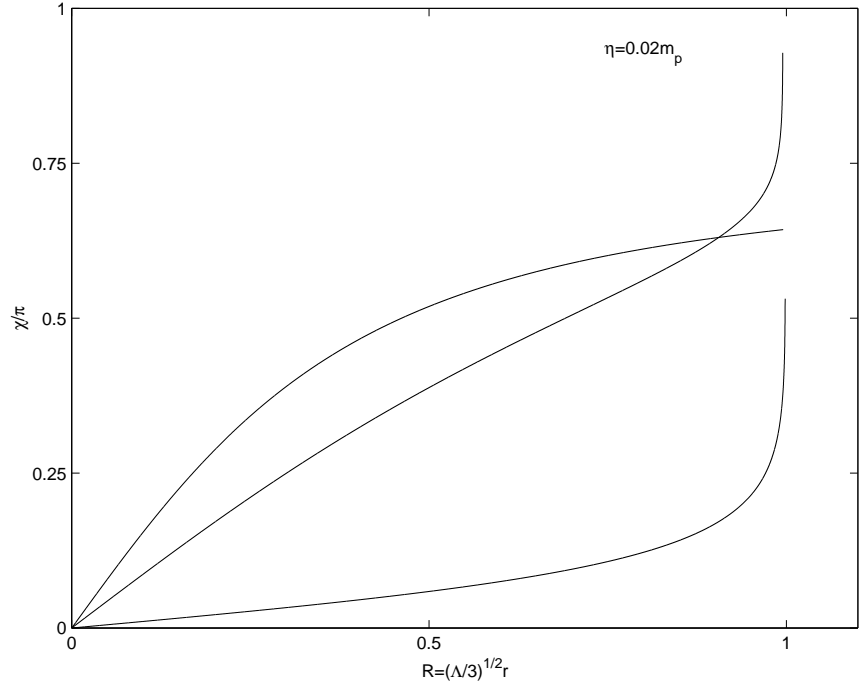,width=4.5in}
\end{center}
\vspace{0.5in}
\caption{
Plot of the self-gravitating scalar field 
for $\eta = 0.02m_p$.
The middle one is the regular solution with
$\chi (R_b)\simeq 0.64295022\pi$.
The others are singular solutions for
$\chi (R_b)=0.9282\pi$ and $\chi (R_b)=0.53149\pi$.}
\label{fig=sfd}
\end{figure}

Let us investigate the regular solution at the horizon.
If we define $Q=1/A$, the scalar field equation~(\ref{eq=scalar}) becomes
\begin{equation}
\chi''+\left( {B'\over 2B}+{Q'\over 2Q}+{2\over R}\right)
-{\sin 2\chi \over QR^2} =0\,.
\label{eq=scalarBQ}
\end{equation}
All physical quantities in the above equation are regular at the horizon.
We perform Taylor expansion for each quantity about the horizon,
$R=R_h+x$, where $B(R_h)=Q(R_h)=0$,
\begin{eqnarray}
B &=& b_1x +b_2x^2 +b_3x^3 + ...\,,\\
Q &=& q_1x +q_2x^2 +q_3x^3 + ...\,,\\
\chi &=& f_0+ f_1x +f_2x^2 +f_3x^3 + ...\,.
\end{eqnarray}
We plug these quantities in the field equation~(\ref{eq=scalarBQ}),
and expand it in the power of $x$.
Equating the coefficient of each term to be zero gives relations 
among coefficients in the expansion series above.
The first term, $1/x$, in the expansion gives
\begin{equation}
f_1-{\sin 2f_0 \over q_1R_h^2}=0\,.
\end{equation}
In other words,
\begin{equation}
\chi'_h +{A_h^2\over A'_h R_h^2}\sin 2\chi_h =0\,.
\label{eq=regBQ}
\end{equation}
Note that this condition reduces to the regularity condition~(\ref{eq=regds})
in de Sitter space for $R_h=1$ and $A=1/(1-R^2)$.
We numerically search the solution which satisfies the above condition
close to the horizon.
We found that the solution exists when $\chi (R_b) \simeq 0.64295022\pi$
for $\eta=0.02m_p$,\footnote{Throughout the paper, we show the numerical
results for $\eta =0.02m_p$ in order to see the gravitational effect clearly.}
where $R_b$ is the location of the last mesh of our numerical calculation.
This is very close to $\chi_h$ for the pure de Sitter solution.
The solution is plotted in Fig.\ref{fig=sfd}.
For this solution $R_b\approx 0.99535$ and $1/A(R_b)\approx 8.7\times 10^{-5}$.
Although we are not able to perform numerical calculations
beyond the horizon, we believe that the configuration is also
very similar to the pure de Sitter one there.

We have also singular solutions similar to the de Sitter ones.
For the singular ones, we performed the calculation up to
$R_b\approx 0.99483$ where $1/A(R_b)\approx 1.0\times 10^{-4}$.
For the texture with one winding number,
we imposed the boundary value of the scalar field, $\chi (R_b)=0.9282\pi$.
Since $\chi$ increases very rapidly close to the horizon, it is reasonable
to impose this boundary value at $R=R_b$ estimated from the de Sitter solution for
$\chi_h=\pi$.
This solution is plotted in Fig.\ref{fig=sfd} as well as
the other singular solution for $\chi (R_b) =0.53149\pi$.
Similar to the de Sitter solutions, there also exist unconventional
solutions for $0.53149\pi < \chi (R_b) \lesssim 0.64\pi$.
This range got narrowed compared with the de Sitter one.

\begin{figure}
\begin{center}
\epsfig{file=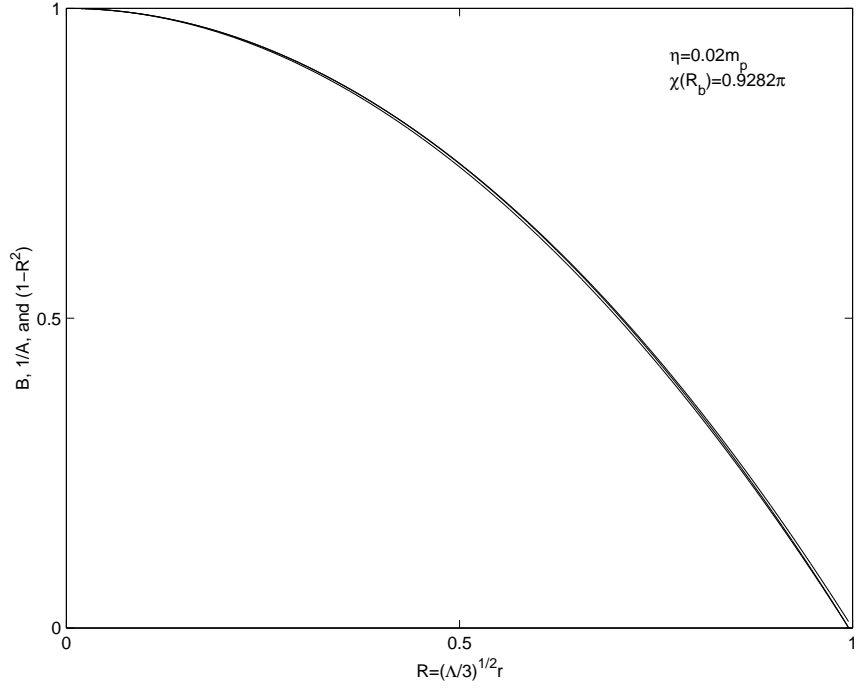,width=4.5in}
\end{center}
\vspace{0.5in}
\caption{
Plot of the gravitational field for $\eta=0.02m_p$
and $\chi (R_b)=0.9282\pi$.
$B$ and $1/A$ are very close to each other
as well as to the pure de Sitter field $1-R^2$.}
\label{fig=AB}
\end{figure}

The gravitational fields of the singular solution for
$\chi (R_b)=-0.9282\pi$ are plotted in 
Fig.\ref{fig=AB}. They are not much different from
the pure de Sitter solutions.

\begin{figure}
\begin{center}
\epsfig{file=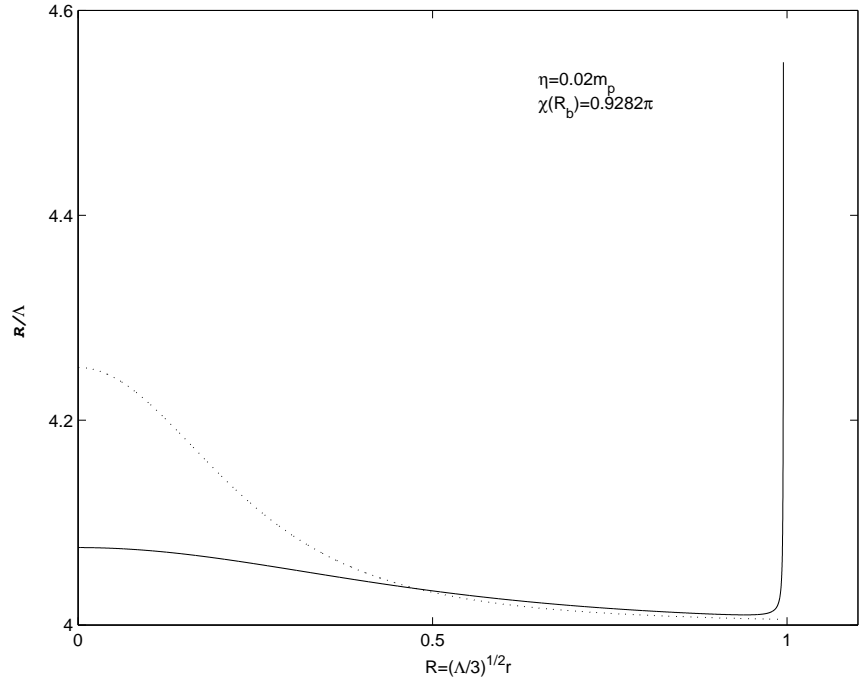,width=4.5in}
\end{center}
\vspace{0.5in}
\caption{
Plot of the Ricci scalar for the singular solution with
$\eta=0.02m_p$ and $\chi (R_b)=0.9282\pi$.
It looks divergent as it approaches the horizon.
The dotted line is for the regular solution.}
\label{fig=ricci}
\end{figure}

As it was discussed in the previous section, the singular
behavior of the derivatives of the scalar field at the horizon
turns the horizon to be singular.
From the regularity condition at the horizon, $\chi''$ is infinite.
We cannot determine the definite value of $\chi'$ there.
It could be either finite, or infinite. 
From our numerical calculations, however, it is obvious that 
$\chi'$ is a very large number.
This indicates that the curvature at the horizon diverges.
The Ricci scalar evaluated from the energy-momentum tensor
depends on $\chi'$, 
\begin{equation}
{\cal R} = 16\pi G\eta^2 \left( {\chi'^2\over 2A}
+{\sin^2\chi \over r^2}\right) + 4\Lambda\,.
\end{equation}
We plot the Ricci scalar for the singular solution in Fig.\ref{fig=ricci}. 
It diverges steeply approaching the horizon.
The other scalar curvatures evaluated from the metric contains
$\chi'$. However, it is not so clear if they involve the terms of 
$\chi''$ which is clearly divergent.

The horizon in pure de Sitter space is an observer-dependent property.
To a given observer $A$, 
it is $A$'s future and past event horizon
beyond which events are causally disconnected with $A$.
Observers at different locations in space have their own
horizon which is different from one another.
If there is a texture centered at $A$'s location,
the spacetime is no longer homogeneous.
In this situation, 
observers have their own horizon which is different from 
the one in pure de Sitter space.
Their horizon is still a local property in terms of causality.
However, the curvature singularity located at $A$'s horizon 
is a global property.
Every observer should recognize this singularity
if it is located inside his/her own horizon.
What makes the horizon, {\it a priori}, a local property of $A$'s
be so special (singular) is that the texture is ``centered''
at $A$'s location.
The way to circumvent the trouble caused by the singularity was
discussed in the previous section.

It is not unusual that a naked singularity exists
in the geometries of topological defects.
For example, global cosmic strings and supermassive gauge
cosmic strings possess a singularity in the spacetime~\cite{Gbstring,Ggstring}.
The difference is that the singularity in the string geometry
is accompanied with the regular scalar field configuration. 
In the case of texture, the singularity arises only when the
scalar field behaves singularly at the horizon;
the horizon is regular if the scalar field is regular there.

\section{Stability}
In previous sections, we found out static texture solutions
co-moving with the expanding background.
In Refs.~\cite{Sornborger,Chen,Barabash}, 
it was investigated that textures are very unstable
even in expanding backgrounds. 
Textures collapse, or expand further. 
It is worthwhile to check if our static solutions are stable, or not.
We adopt the linear-perturbation method used, 
for example, in Ref.~\cite{Zhou} where
authors check the stability of black holes.
This method finds if there is an unstable mode in perturbations, but 
does not describe the evolution precisely after the perturbations.

Let us expand the gravitational fields and the scalar field as
\begin{eqnarray}
B(t,r) &=& B_0(r) +\epsilon B_1(t,r) + {\cal O}(\epsilon^2)\,,
\label{eq=expansionB}\\
A(t,r) &=& A_0(r) +\epsilon A_1(t,r) + {\cal O}(\epsilon^2)\,,
\label{eq=expansionA}\\
\chi(t,r) &=& \chi_0(r) +\epsilon \chi_1(t,r) + {\cal O}(\epsilon^2)\,,
\label{eq=expansionchi}
\end{eqnarray}
where $B_0(r)$, $A_0(r)$, $\chi_0(r)$ are the static solutions we obtained
in the previous section, and $\epsilon$ is a small number.
We expand the time-dependent field equations in the power of $\epsilon$,
and keep the terms only up to the first-order in $\epsilon$.
The time-dependent Einstein equations are
\begin{eqnarray}
-G^t_t &=& {1\over A}\left( {A'\over Ar}+ {A-1\over r^2}\right)
	=8\pi G\eta^2 \left( {\dot{\chi}^2 \over 2B}
	+{\chi'^2\over 2A}+{\sin^2\chi \over r^2}\right)
	+\Lambda\,,\\
-G^r_r &=& {1\over A}\left( -{B'\over Br}+ {A-1\over r^2}\right)
	=8\pi G\eta^2 \left( -{\dot{\chi}^2 \over 2B}
	-{\chi'^2\over 2A}+{\sin^2\chi \over r^2}\right)
	+\Lambda\,,\\
G_{tr} &=& {\dot{A}\over Ar} = 8\pi G\eta^2\dot{\chi}\chi'\,,
\end{eqnarray}
and the scalar field equation is
\begin{equation}
{\ddot{\chi} \over B}-{1\over B}\left( {\dot{B}\over 2B}
-{\dot{A}\over 2A}\right)\dot{\chi}
-{\chi''\over A} -{1\over A}\left({B'\over 2B}-{A'\over 2A}+{2\over r}
\right)\chi' +{\sin 2\chi \over r^2}=0\,.
\end{equation}
We plug expansions~(\ref{eq=expansionB})-(\ref{eq=expansionchi}) 
in the above field equations.
The first-order terms in $\epsilon$ gives equations as following.

From the $G_{tr}$ equation,
\begin{equation}
{\dot{A_1} \over A_0r} = 8\pi G\eta^2\chi_0'\dot{\chi_1}\,,
\end{equation}
of which solution is
\begin{equation}
{A_1 \over A_0r} = 8\pi G\eta^2\chi_0'\chi_1 + f(r)\,,
\label{eq=A1}
\end{equation}
where $f(r)$ is an integration constant w.r.t. time.
This solution satisfies $G^t_t$ equation and
we can fix the integration constant, $f(r)=0$, as a boundary condition.
This gives the unique solution to Eq.~(\ref{eq=A1}).
This boundary condition is appropriate because it is
consistent with reasonable initial conditions,
$A(t=0,r)=A_0(r)$ and $\chi (t=0,r)=\chi_0(r)$. 
This means that perturbations are zero initially,
$A_1(0,r)=0$ and $\chi_1 (0,r)=0$, but their velocities are not.

The $G^t_t+G^r_r$ equation gives
\begin{equation}
{1\over A_0}\left( {B_1'\over B_0}-{A_1'\over A_0}
-{B_0'B_1\over B_0^2}+{A_0'A_1\over A_0^2}\right)
={A_1\over A_0^2} \left( {B_0'\over B_0}
-{A_0'\over A_0}+{2\over r}\right)
-16\pi G\eta^2 {\sin 2\chi_0 \over r}\chi_1\,.
\label{eq=B1}
\end{equation}
The scalar field equation gives
\begin{eqnarray}
{\ddot{\chi_1} \over B_0} &-& {\chi_1''\over A_0}
-{1\over A_0}\left( {B_0'\over 2B_0}-{A_0'\over 2A_0}
+{2\over r}\right)\chi_1'
+{2\cos 2\chi_0\over r^2}\chi_1 \nonumber\\
{}&-& {1\over 2A_0}\left( {B_1'\over B_0}-{A_1'\over A_0}
-{B_0'B_1\over B_0^2}+{A_0'A_1 \over A_0^2}\right)\chi_0'
+{A_1\over A_0}{\sin 2\chi_0\over r^2} =0\,.
\label{eq=chi1}
\end{eqnarray}
Using Eqs.~(\ref{eq=A1}) and (\ref{eq=B1}),
the perturbation terms $B_1$, $A_1$, $B_1'$, and $A_1'$ are
nicely eliminated in the scalar field equation~(\ref{eq=chi1}),
\begin{equation}
{A_0\over B_0}\ddot{\chi_1} -\chi_1''
+ a(r)\chi_1'
+ b(r)\chi_1 =0\,,
\label{eq=chi}
\end{equation}
where
\begin{eqnarray}
a(r) &=& -{B_0'\over 2B_0}+{A_0'\over 2A_0}
-{2\over r}\,,\\
b(r) &=& A_0 \left[ {2\cos 2\chi_0 \over r^2}
+8\pi G\eta^2\chi_0'r \left[ {2\sin 2\chi_0 \over r^2}
-{1\over A_0}\left( {B_0'\over 2B_0}-{A_0'\over 2A_0}
+{1\over r}\right)\chi_0'\right]\right]\,.
\end{eqnarray}
Therefore, the perturbation equation for $\chi_1$
involves only the quantities obtained from static solutions.
Let us take the perturbed scalar field in the form,
\begin{equation}
\chi_1(t,r) =\varphi (r)e^{i\omega t}\,,
\end{equation}
then the equation for the amplitude $\varphi (r)$ becomes
\begin{equation}
{B_0\over A_0}[-\varphi''(r) +a(r)\varphi'(r)+b(r)\varphi (r)]
=\omega^2\varphi (r)\,.
\label{eq=varphi}
\end{equation}
This is an eigenvalue equation with the eigenvalue $\omega^2$.
We rescale the radial coordinate and amplitude function by
\begin{eqnarray}
z &=& \int^r_0\sqrt{A_0\over 2B_0} dr\,,\\
\psi (z) &=& N {\varphi (r)r \over z}\,,
\end{eqnarray}
where $N$ is a normalization constant.
This rescaling puts the equation into the non-relativistic
Schr\"odinger-type equation in spherical coordinates
\begin{equation}
\left[ -{1\over 2}{d^2\over dz^2}-{1\over z}{d\over dz} +V(z)
\right]\psi (z) =\omega^2 \psi (z)\,,
\label{eq=psi}
\end{equation}
where the potential is 
\begin{equation}
V[z(r)] = {B_0\over A_0}\left[ {1\over 2r}
\left({B_0'\over B_0}-{A'_0\over A_0}\right) +b(r)\right]\,.
\end{equation}
The advantage of deriving the equation into this form is
that the shape of the potential $V$ provides insight for
the eigenvalue spectrum.
If $V$ is a potential well, there exists a discrete
eigenvalue spectrum.
If the minimum of $V$ is negative, there possibly exists
negative eigenvalues, so $\omega$ can be imaginary. 
In such a case, the perturbed field has an exponentially
increasing, or decreasing solution depending on the
boundary condition we impose.
Even if there exists only one negative eigenvalue of $\omega^2$,
the system becomes completely unstable.

\begin{figure}
\begin{center}
\epsfig{file=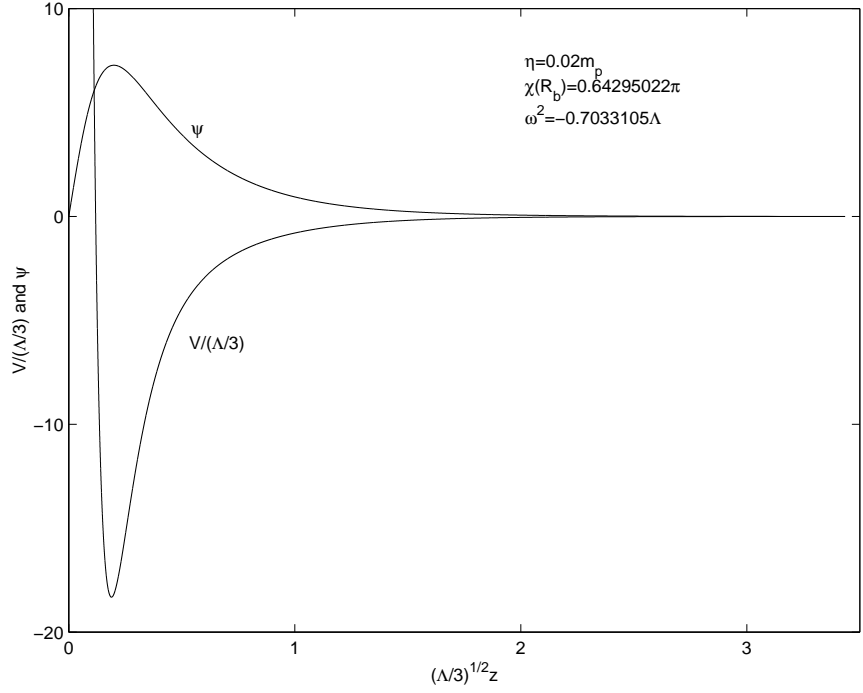,width=4.5in}
\end{center}
\vspace{0.5in}
\caption{
Plot of the potential $V$ for the perturbation field
of the regular solution with $\eta =0.02m_p$.
There exists an unstable mode with the eigenvalue
$\omega^2= -0.7033105\Lambda$.
The corresponding eigenfunction $\psi$ is also
plotted. 
The normalization of $\psi$ was not taken into account.}
\label{fig=Vreg}
\end{figure}

Now with the given equation above, let us examine the
stability of the solutions obtained in the previous section.
First, for the regular solution, the potential $V$ is plotted
in Fig.\ref{fig=Vreg}.
It is a potential well with a negative minimum.
We numerically solve the eigenvalue equation~(\ref{eq=varphi})
rather than Eq.~(\ref{eq=psi}).
As the potential $V$ diverges at origin, 
we set the boundary condition, $\psi (0)=0$, so
$\varphi (0)=0$. For the regular solution at the horizon,
$\chi_0'(r_h)$ is finite, and $B_0(r_h)=1/A(r_h)=0$.
Therefore, it is easy to see $\varphi (r_h) =0$.
With these two boundary conditions for $\varphi (r)$,
we found a negative eigenvalue, $\omega^2 = -0.7033105\Lambda$.
The corresponding eigenfunction $\psi (r)$ is plotted in 
Fig.\ref{fig=Vreg}.
Our regular solution is unstable under small perturbations.

\begin{figure}
\begin{center}
\epsfig{file=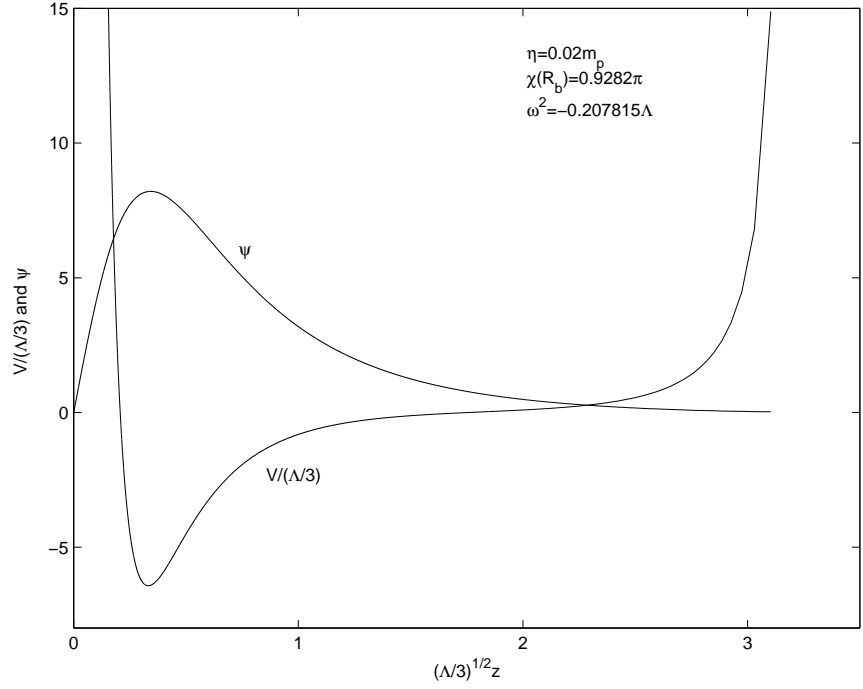,width=4.5in}
\end{center}
\vspace{0.5in}
\caption{
Plot of the potential $V$ for the perturbation field
of the singular solution with 
$\eta=0.02m_p$ and $\chi (R_b)=0.9282\pi$.
There exists an unstable mode with the eigenvalue
$\omega^2= -0.207815\Lambda$.
The corresponding eigenfunction $\psi$ is also
plotted. 
The normalization of $\psi$ was not taken into account.}
\label{fig=Vsing}
\end{figure}

Now, let us examine singular solutions.
We are particularly interested in the solution with $\chi_0(r_h)\sim\pi$,
which completes one winding in the region under consideration.
The potential $V$ is plotted in Fig.\ref{fig=Vsing}.
The potential diverges at the origin similar to that of the regular one.
And, as we approach the horizon, $r\to r_h$ ($z\to\infty$),\footnote{Even though
finite $z$ region is shown in the plot, $z$ goes to infinity as we approach
the horizon. This is evident if we consider the pure de Sitter case where
analytic form of radial rescale is possibly obtained,
$z=(r_h/\sqrt{2})\tanh^{-1}(\sqrt{r/r_h})$.
We insist that the situation be similar in the self-gravitating case.}
the potential also diverges.
Therefore, we impose boundary conditions, $\psi (0) = \psi (z\to\infty)=0$,
which, in turn, gives $\varphi (0) = \varphi (r_h)=0$.
Numerical calculation shows that there exists a negative
eigenvalue, $\omega^2 =-0.207815\Lambda $. The corresponding eigenfunction
is plotted in Fig.\ref{fig=Vsing}.
The static singular solution with $\chi_0(r_h)\sim\pi$ is unstable.

For the other singular solutions with 
$0< \chi_0(r_b) \lesssim 0.53149\pi$, 
the potential $V$ has a negative minimum also.
However, it is very close to zero. There does not necessarily
exist a negative eigenvalue in such a case. 
In particular, if the symmetry-breaking
scale $\eta$ is raised, the minimum value of $V$ is raised above zero.
Therefore, we conclude that this type of singular solutions is stable.

Even though we are not so convinced to study the unconventional solutions
with $0.53149\pi < \chi_0(r_b) \lesssim 0.64\pi$,
the potential $V$ for those solutions has a deep negative minimum,
and the solutions are unstable.
Therefore, in the rest of this paper,
we will consider that textures are stable only for 
$\chi_b \lesssim 0.53149\pi$.

As a result, the static regular solution and singular solutions with
a high winding number up to one (or, a little bit higher than one) 
are unstable under perturbations.\footnote{The winding number is
evaluated as $W = \chi_b/\pi - (\sin 2\chi_b)/2\pi$,
where $\chi_b$ is the value of the scalar field at the boundary
under consideration.
In this work, however, we shall mostly use $\chi_b$ for numerical values  
rather than elaborating to calculate the corresponding $W$.}
Depending on the boundary conditions imposed at the moment of
perturbations, textures collapse, or expand w.r.t. the 
expanding background.



The perturbed expansion proceeds behind the cosmological
horizon, so it is not very interesting cosmologically.
The collapse will be discussed in the next section.

\section{Cosmological evolution}
In this section, we qualitatively discuss the evolution of the
configurations we obtained.
As we mentioned earlier, the configurations are self-similar
regardless of the $\Lambda$ scale.
Imposing a definite $\Lambda$ value fixes the physical length scale.

If $\Lambda$ is very large and dominates the energy density of the universe,
the situation corresponds to inflation.
The large $\Lambda$ plays the role of 
the vacuum energy given by the inflaton field.
In this case, the cosmological horizon scale determined by the total energy
of the universe is comparable to the horizon scale $r_h \sim (\Lambda /3)^{-1/2}$  
of our solution.
Then the whole configuration we obtained up to the horizon is applicable.

Unstable textures collapse during inflation.
Even if the collapse reaches a relativistic speed, however,
it would not be able to compensate the inflationary expansion.
Textures survive for the inflation era.
After inflation ends, textures will collapse to the origin very quickly,
and unwind themselves in the end.

For the rest of the history of the universe except present, 
$\Lambda$ remains very small compared with the total energy.
The cosmological horizon scale is much smaller than $r_h$.
If textures are formed at this stage,
the causality bound prohibits the formation of a large texture extended
to $r_h$. Neighboring textures are formed before that.
Even though such a large texture had been formed 
with a high winding number up to one,
the observers would see only a part of it, so
an incomplete winding in the universe.
After textures start to collapse, they will see the scalar field
with higher values crossing into the boundary of the universe.
In the end, the observers who initially saw a part of the winding, will
see the complete winding of the field.

We found that the static textures for $\chi_b \gtrsim 0.53149\pi$
are unstable.
Even though the universe with a cosmological constant is topologically
different from those without it, (for example, there is no static solution
for high winding numbers in the others) collapsing behavior is very similar.
In Ref.~\cite{Sornborger}, it was investigated that textures collapse for
winding number $W \geq W_c \sim 0.6$ in the flat, radiation, and matter background.
Depending on the initial configuration imposed,
the collapsing time varies.
It can be explained as following. 
For $W \geq W_c$, there does not exist a static solution in such backgrounds.
Imposing the initial configuration already implies a certain amount of perturbation.
The higher $W$ is, the larger are the perturbations.
For our case, the static configuration will start to collapse slowly.
However, the collapse will reach a relativistic speed in the end.

Textures with $\chi_b \lesssim 0.53149\pi$ are stable according to our investigation.
As it was pointed out in Ref.~\cite{Sornborger}, they behave as a
local fluctuation, and will decay into Goldstone bosons.

\section{Conclusions}
In this work, we investigated O(4) textures with a positive 
cosmological constant. It has been previously investigated that
O(4) textures collapse in the flat, self-gravitating, radiation, 
and matter background.
There was no static solution found out in those backgrounds,
especially for high winding number textures. 
However, with a cosmological constant, we found out that
O(4) textures have static configurations
for all winding numbers up to slightly higher than one.
De Sitter expansion of the background prevents the texture from collapsing.
However, the static configuration is unstable under small perturbations
if the winding number is larger than some critical value, 
$\chi_b \gtrsim 0.53149\pi$ (for $\eta =0.02m_p$).

The textures with such high winding numbers will collapse 
after the perturbations are applied.
Initially the collapse will occur slowly, but it will reach a relativistic
speed in the end. At the final stage, the field gradient energy
will be concentrated about the center, and textures will unwind.
Meanwhile, textures will leave a signature in the large-scale structure
formation of the universe. 
The resulting signature will be similar to that of
the textures in the other backgrounds.

For $\chi_h \lesssim 0.53149\pi$, textures are stable.
They will decay into Goldstone bosons in the end.

The static configurations we obtained fit the region inside the horizon.
They co-move with the expanding background while they are stable.
The scalar field is definite everywhere within the horizon.
For most of the configurations except the regular one,
the static solutions have singular field derivatives at the horizon.
This singular behavior turns the horizon to become a curvature singularity.
However, this is not a matter so much because the horizon is always
located outside the cosmological causal bound.

The detailed evolution based on solving time-dependent field equations
was not investigated in this work.
It will be interesting to perform numerical simulations for such a evolution.
In this case, we can extend the model to a quintessence field background
of which equation of state is $P = w\rho$ ($-1<w<0$).
For such a field, there would not exist a static description
of the spacetime. Therefore, numerical simulation is very useful to investigate
the evolution. 
The quintessence field may affect the texture evolution more than the cosmological
constant, because its energy density could be relatively high at the early
stage of the universe.

\acknowledgements
I am grateful to Chryssomalis Chryssomalakos, Jemal Guven, Hernando Quevedo,
Marcelo Salgado, and Daniel Sudarsky for helpful discussions.
I also thank to Tanmay Vachaspati for encouragement and discussions.
This work was supported in part by
Proyecto de Investigacion DGAPA IN119799.

\clearpage

\clearpage

\clearpage

\clearpage

\clearpage

\clearpage

\end{document}